\documentclass[aps,prl,twocolumn,groupedaddress,amssymb,showpacs]{revtex4}
\usepackage[dvips]{graphicx}
\usepackage{ulem}
\begin{document}

\preprint{APS/123-QED}

\title{Hyperfine frequency shift in two-dimensional atomic hydrogen}

\author{A.\,I. Safonov$^1$}
\email{safonov@isssph.kiae.ru}
\author{I.\,I. Safonova$^1$}
\author{I.\,S. Yasnikov$^2$}
\affiliation{
$^1$Russian Research Centre Kurchatov Institute, pl. Kurchatova 1, Moscow, 123182 Russia\\
$^2$Togliatti State University, Togliatti, 445667, Russia}

\date{\today}

\begin{abstract}
We propose the explanation of a surprisingly small hyperfine frequency shift in the
two-dimensional (2D) atomic hydrogen bound to the surface of superfluid helium below 0.1 K. Owing
to the symmetry considerations, the microwave-induced triplet-singlet transitions of atomic pairs
in the fully spin-polarized sample are forbidden. The apparent nonzero shift is associated with
the density-dependent wall shift of the hyperfine constant and the pressure shift due to the
presence of H atoms in the hyperfine state $a$ not involved in the observed $b\rightarrow c$
transition. The interaction of adsorbed atoms with one another effectively decreases the binding
energy and, consequently, the wall shift by the amount proportional to their density. The pressure
shift of the $b\rightarrow c$ resonance comes from the fact that the impurity $a$-state atoms
interact differently with the initial $b$-state and final $c$-state atoms and is also linear in
density. The net effect of the two contributions, both specific for 2D hydrogen, is comparable
with the experimental observation. To our knowledge, this is the first mentioning of the
density-dependent wall shift. We also show that the difference between the triplet and singlet
scattering lengths of H atoms, $a_t-a_s=30(5)$~pm, is exactly twice smaller than the value
reported by Ahokas {\it et al.}, Phys. Rev. Lett. {\bf101}, 263003 (2008).
\end{abstract}

\maketitle

Hyperfine atomic transitions, which are currently used as frequency standards~\cite{freq_std} for
their extreme stability, experience a frequency shift due to the interaction of resonant atoms
with the container walls, buffer gas and with each other. These are commonly referred to as the
wall shift, pressure shift and contact shift, respectively. The contact or collision shift is
associated with a difference in the scattering length of atoms in two hyperfine states coupled by
the resonance. It has been shown quite convincingly, both in theory and experiments with
ultra-cold alkali vapors, namely, with bosonic $^{87}$Rb~\cite{Harber} and fermionic
$^6$Li~\cite{Zwierlein}, that the measured resonance shift due to the interstate collisions is
independent of the coherence and polarization in the two-level system and fully determined by the
statistics of atoms: the magnitude of the shift is the equilibrium energy splitting between the
two internal states in a fully decohered cloud multiplied by the two-particle correlation
function, which is $g^{(2)}=1$ for distinguishable particles and Bose-Einstein condensates,
$g^{(2)}=2$ for thermal bosons and $g^{(2)}=0$ for fermions~\cite{Zwierlein}.

It is especially surprising in this context that the hyperfine frequency shift recently observed
in the 2D Bose gas of spin-polarized atomic hydrogen (H${\downarrow}$) appeared to be a factor of
$\sim120$ less than expected~\cite{Ahokas}. This contradiction, i.e., the virtual absence of the
contact shift and the respective ``fermionic'' behavior of atomic hydrogen, has been lately
resolved by taking into account a symmetry selection rule that forbids the electronic
triplet-singlet transitions of atomic pairs: under the absorption of microwave quanta, the
electron spins of all atoms experience coherent rotation being parallel to each other so that the
interaction energy does not change~\cite{Safonov_08}. However, the nature of the small yet nonzero
hyperfine frequency shift in atomic hydrogen remained unclear. The question was lately addressed
by Hazzard and Mueller~\cite{Hazzard} and Ahokas {\it et al.}~\cite{Ahokas_3D}. However, as we
show below, their arguments are contradictory at some points and incomplete.

In this Letter, we propose the alternative explanation of the apparent frequency shift in 2D
atomic hydrogen based on the independent action of two different factors, the density-dependent
wall shift and the pressure shift due to the residual atoms in the hyperfine state that is not
involved in the observed resonance and has the opposite projection of a nuclear spin. We also show
that the difference between the triplet and singlet scattering lengths of H atoms,
$a_t-a_s=30(5)$~pm, is exactly twice smaller than the value reported in Ref.~\cite{Ahokas_3D}.

The ground state of a H atom in a magnetic field $B$ is split into four hyperfine states (in the
$\vert m_s,\,m_i\rangle$ basis in the order of increasing energy):
\begin{eqnarray}
&a=\cos\theta\vert-+\rangle-\sin\theta\vert+-\rangle, &b=\vert--\rangle,\nonumber\\
&c=\cos\theta\vert+-\rangle+\sin\theta\vert-+\rangle, &d=\vert++\rangle,\nonumber
\end{eqnarray}
where $\pm$ stands for $\pm\frac{1}{2}$, $\tan(2\theta)=A[(\gamma_e+\gamma_p)hB]^{-1}$,
$\gamma_e$($\gamma_p$) is the electron (proton) gyromagnetic ratio and $A/h=1420$~MHz is the
hyperfine constant of hydrogen. The experiments~\cite{Ahokas} with 2D hydrogen were carried out at
$T\sim0.1$~K in the high field $B=4.6$~T such that $\gamma_ehB\gg A,\,k_BT$ and therefore the
electron spins of atoms are polarized. Moreover, the sample is almost entirely in the
doubly-polarized state $b$. The resonance line of the electron spin-flip $b\rightarrow c$ Zeeman
transition of the atoms adsorbed on a liquid helium surface is shifted with respect to the
resonance of the gas-phase atoms by the dipole field $B_d$ imposed by spin-aligned atoms on each
other~\cite{Instability}. To exclude the unwanted dipole contribution, Ahokas {\it et al.} used
another, nuclear spin-flip, $b\rightarrow a$ transition as a reference. The latter was supposed to
be free from the contact shift.

In the high-field ($\gamma_e\hbar B\gg A$) approximation, the frequency difference of the
$b\rightarrow a$ resonance on the surface and in the bulk gas is~\cite{Ahokas}
\begin{equation}\label{Delta_nu_fin}
\Delta\nu_{ab}\equiv\nu_{ab}^s-\nu_{ab}^0=\frac{\Delta
A_w}{2h}\left(1+\frac{\gamma_p}{\gamma_e}\right)-\frac{\gamma_p}{\gamma_e}\Delta\nu_c.
\end{equation}
Here $\Delta A_w$ is a change in the hyperfine constant due to the interaction with liquid helium
and $\Delta\nu_c$ is the expected contact shift of the $b\rightarrow c$ resonance. In fact,
$\Delta A_w$ also known as the wall shift is proportional to a change in the unpaired electron
density at the proton due to the distortion of the electron wavefunction of the atom in the
adsorption potential. $\gamma_p/\gamma_e\approx1.5\times10^{-3}$ may be safely neglected with
respect to unity. Note, that the $b\rightarrow a$ transitions on the surface and in the bulk gas
are excited in different external fields, such that the frequency of the respective $b\rightarrow
c$ resonance does not change.

Experimentally, $\Delta\nu_{ab}$ is linear in the surface density $\sigma_b$ of atoms in the
hyperfine state $b$
\begin{equation}\label{Delta_nu-exp}
\Delta\nu_{ab}=C_0+C_1\sigma
\end{equation}
with $C_0=-24.79(2)$~kHz and $C_1=1.52(15)\times10^{-9}$~Hz$\cdot$cm$^2$~\cite{Ahokas}. Ahokas
{\it et al.} attributed this entirely to the density variation of the contact shift $\Delta\nu_c$.
Thus, the values actually quoted in Ref.~\cite{Ahokas} were those derived for $\Delta A_w$ and
$\Delta\nu_c$. They further argued that, like in alkalies, $\Delta\nu_c$ should be proportional to
a difference $a_t-a_s$ between the triplet $b-b$ and singlet $b-c$ scattering length in the
initial and final state, respectively. Owing to a small value of $C_1$ ($\sim120$ times less than
expected), Ahokas {\it et al.} concluded that the scattering lengths are probably considerably
different from the common values, $a_t=0.72$ and $a_s=0.17$~\AA~\cite{Williams}.

Shortly after that, Safonov {\it et al.}~\cite{Safonov_08} showed that, due to the symmetry
constraints, both electronic and nuclear triplet-singlet transitions of atomic pairs are strictly
forbidden in the spatially uniform sample hundred-percent polarized in both electron and nuclear
spins. In fact, scattering of two atoms is determined by their total spin $F$, whose parity gives
the parity of the relative angular momentum, i.e., the possibility of the $s$-wave scattering, and
the total electron spin $S$, which determines the interaction potential (singlet or triplet) and
therefore the scattering length. The initial spin wavefunction of two $b$-state atoms ($F=2$,
$m_F=-2$; $S=I=1$, $m_S=m_I=-1$) is \textit{symmetric} with respect to the permutation of
particles. The same holds for their unperturbed Hamiltonian and the perturbation Hamiltonian due
to the microwave field. Consequently, the final state after the absorption of a microwave photon
$\vert bc\rangle_g=\frac{1}{\sqrt{2}}(\vert bc\rangle+\vert cb\rangle)$ is also
\textit{symmetric}. Thus, the total spin $F$ must remain even. The total electron and nuclear
spins are necessarily conserved: $S=I=1$. Notably, this holds in an arbitrary magnetic field,
which is easily verified by evaluating $\langle bc\vert\hat{S}^2\vert bc\rangle_g$ and $\langle
bc\vert\hat{I}^2\vert bc\rangle_g$. In a classical interpretation, all electron spins of the
ensemble of hydrogen atoms are coherently tilted so that each atomic pair constitutes an
electronic triplet, just like in the initial state. Thus, the contact shift $\Delta\nu_c$ must be
exactly zero irrespective of the actual values of the scattering lengths $a_t$ and $a_s$. The
revision of $a_t$ and $a_s$ is therefore unnecessary. The situation with the nuclear $b\rightarrow
a$ Zeeman transition is obviously identical.

Recently, Hazzard and Mueller~\cite{Hazzard} tried to explain the smallness of the hyperfine shift
in the 2D hydrogen in a different way. They suggested that the substrate-mediated H-H interaction
is substantially different from the interaction of free atoms and concluded that the hyperfine
frequency shift of the $b\rightarrow c$ transition must be much {\it lower} in the adsorbed phase,
as compared to the bulk gas. However, they quite arbitrarily ignored the above symmetry arguments,
which imply that the contact shift of the $b\rightarrow c$ and $b\rightarrow a$ transitions is
exactly zero in both 2D and 3D cases, as long as the gas is doubly spin-polarized. This makes the
theory of Hazzard and Mueller irrelevant in this respect. Note, that Ahokas {\it et
al.}~\cite{Ahokas_3D}, who basically repeated the arguments of Ref.~\cite{Safonov_08} regarding
their measurements of the contact shift in the 3D atomic hydrogen, also came to a conclusion that
the shift of the $b\rightarrow c$ transition is zero in the doubly polarized gas. In addition, the
conclusion of Ref.~\cite{Hazzard} contradicts with the observation~\cite{Ahokas, Ahokas_3D} that
the contact shift in the 2D gas is {\it higher} than in the bulk phase, as follows from comparison
of the respective experiments.

Safonov {\it et al.}~\cite{Safonov_08} discussed the opportunity that the frequency shift of the
$b\rightarrow c$ transition may be nonzero due to the dipole-dipole interaction of electron spins.
It was asserted that the dipole-dipole interaction is symmetric with respect to the permutation of
atoms and therefore does not break the symmetry constraint for the hyperfine transitions of atomic
pairs. However, Ahokas {\it et al.}~\cite{Ahokas_3D} attributed the nonzero shift to the dipole
mechanism saying that the dipole-dipole interaction is long-range and must be treated separately.
Unfortunately, in their brief discussion, they did not show how the symmetry constraint is lifted
by the dipole-dipole interaction. Neither they explained the difference between this mechanism and
the usual dipole shift of the hyperfine resonance due to the internal magnetic field $B_d$ imposed
by spin-polarized atoms on each other. In this regard, it is crucial to show that this effect acts
differently on the $b\rightarrow c$ and $b\rightarrow a$ transitions, otherwise it cannot be
detected by the technique of Ref.~\cite{Ahokas_3D}. It is also unclear how all this applies to the
2D case. Ref.~\cite{Ahokas_3D} refers to the future publication of the calculation details, which
does not allow a more specific discussion of the subject.

However, there is another reason for $\Delta\nu_c$ to be nonzero in hydrogen. This is due to the
residual atoms in the hyperfine state $a$ that are always present in the sample owing to
spontaneous one- and two-body nuclear relaxation followed by relatively fast recombination with
$b$ atoms into H$_2$ molecules. The role of these third-state atoms in the 3D case has been
demonstrated in recent experiments of Ahokas {\it et al.}~\cite{Ahokas_3D}. However, their results
for the magnitude of the contact shift $\Delta\omega_{ad}$ of the $a\rightarrow d$ transition in
the $b$-state gas and the interaction energy $E_{ab}$ of $a$ and $b$-state atoms are exactly
one-half of the respective values obtained in Ref.~\cite{Safonov_08}. We believe that this
difference in $\Delta\omega_{ad}$ and $E_{ab}$ originates from the fact that, in the calculation
of the interaction energy, the authors of Ref.~\cite{Ahokas_3D} considered hydrogen atoms in
different hyperfine states as {\it distinguishable} and used $g^{(2)}=1$. At the same time, they
expressed the diatomic states in terms of electronic and nuclear singlets and triplets, which
assumes {\it identical} particles with $g^{(2)}=2$ and includes their ``distinguishability''
automatically, via the ratio of symmetric and antisymmetric contributions to the diatomic
wavefunction. As a result, the difference in the scattering length quoted in Ref.~\cite{Ahokas_3D}
must be divided by 2, which yields $a_t-a_s=30(5)$~pm. This is somewhat lower than the theoretical
values~\cite{Williams}.

The fraction of $a$ atoms is typically very small and difficult to detect by ESR. Therefore, we
neglect the interaction of these atoms with each other. The interaction energies of $ac$ and $bd$
pairs are essentially the same~\cite{Safonov_08}. Thus, the shift of the $b\rightarrow c$
resonance in the presence of $a$ atoms is (cp. Eq. (2) of Ref.~\cite{Ahokas_3D} with the
above-said factror-of-two correction and Eq. (8) of Ref.~\cite{Safonov_08})
\begin{equation}\label{Delta_nu_bc}
\Delta E_\textrm{int}\equiv h\Delta\nu_{bc}=\frac{2\pi\hbar^2}{ml}(a_s-a_t)\sigma_a,
\end{equation}
where $\sigma_a$ is the 2D density of $a$ atoms and $l\sim5$\AA\, is the delocalization length of
adsorbed atoms' wavefunction in the surface-normal direction.

Let us now consider the rate equations including relaxation and recombination to evaluate the
steady-state density of $a$ atoms. In the experiments under consideration, the great majority of
hydrogen atoms, predominantly in the $b$ state, is in the gas phase in the sample cell volume. On
the other hand, nearly all recombination and relaxation events occur in the 2D phase adsorbed on
either the low-density ``warm'' ($T_0=120-200$ mK) cell walls or on the high-density cold spot
($T_s=50-90$ mK)~\cite{Jarvinen,Ahokas}. Thus, the total number of $a$ atoms in the cell obeys the
rate equation
\begin{eqnarray}\label{Na}
\dot{N}_a=\Phi_a+A(G_1\sigma_b+G_2\sigma_b^2-K_{ab}\sigma_a\sigma_b)+\nonumber\\+
A_s(G_{1s}\sigma_{bs}+G_{2s}\sigma_{bs}^2-K_{abs}\sigma_{as}\sigma_{bs}).
\end{eqnarray}
Here $\Phi_a$ is the flux of atoms from the low-temperature dissociator/polarizer ~\cite{Ahokas}
and $G_1$,  $G_2$ and $K_{abs}$ are the rate constants of, respectively, one- and two-body nuclear
relaxation and exchange recombination on the cell walls. The subscript $s$ refers to the values on
the cold spot. The area of the cell walls and the cold spot is $A\sim100$ cm$^2$ and $A_s=0.32$
cm$^2$, respectively.
%In the steady state ($\dot{N}_a = 0$), assuming $\Phi_a = 0$ we obtain
%\begin{eqnarray}\label{as_stat}
%\left(\frac{A}{A_s}\right)\left(\frac{\sigma_b}{\sigma_{bs}}\right)(G_1+G_2\sigma_b-K_{ab}\sigma_a)+\nonumber\\
%+ (G_{1s}+G_{2s}\sigma_{bs}-K_{abs}\sigma_{as}) = 0
%\end{eqnarray}
The ratio of the equilibrium $b$-atom density on the cold spot to that on the cell walls is
$\sigma_{bs}/\sigma_b\approx\exp[E_a(T_s^{-1}-T_0^{-1})]$, which  amounts to about 3$\times10^4$
at $T_0$ = 200 mK and $T_s$ = 70 mK. Therefore, $\alpha\equiv A\sigma_b/A_s\sigma_{bs}\sim0.1$. In
this case, the two-body relaxation on the cell walls may be safely omitted. In addition, it is
reasonable to assume that $\sigma_a\ll\sigma_{as}$. Then, in the steady state ($\dot{N}_a = 0$),
neglecting the incoming flux $\Phi_a$ we obtain
\begin{equation}\label{as}
\sigma_{as}\approx\frac{G_{1s}+\alpha G_1}{K_{abs}}+\frac{G_{2s}}{K_{abs}}\,\sigma_{bs}.
\end{equation}
Thus, the density of $a$ atoms on the cold spot is linear in the density of $b$ atoms. In
experiment, $\dot{N}_b=0$ is maintained. This implies that the flux of $b$ atoms replenishes the
recombination losses of atoms in both hyperfine states (they recombine in equal numbers).
Obviously, due to the operation of the polarizer, $\Phi_a\ll\Phi_b$. Therefore, the smallness of
$\Phi_a$ in Eq. (\ref{Na}) is verified.

The rate constant of the first-order nuclear relaxation on the cell walls is $G_1\sim0.1$
s$^{-1}$~\cite{Ahokas}. $G_{1s}$ is probably close to $G_1$. According to Safonov {\it et
al.}~\cite{OurEa}, $K_{ab}=2.8\times10^{-9}\cdot T^{3/2}$
cm$^2$K$^{-3/2}$s$^{-1}\simeq5\times10^{-11}$ cm$^2$s$^{-1}$ at 70 mK. In theory, the two-body
relaxation rate constant is zero if the surface is exactly normal to the ambient field~\cite{Ahn}.
However, this was never observed in experiment. The experimental estimate for the field of
interest obtained from the multi-parameter fits of the decay curves is
$G_{2s}\lesssim4\times10^{-13}$ cm$^2$/s~\cite{Bell}. The theoretical value averaged over possible
local orientations of the surface with respect to the ambient field (due to the surface roughness)
is $G_{2s}\approx1.4\times10^{-13}$ cm$^2$/s in $B=4.6$ T and essentially temperature-independent.
This yields  the slope $d\sigma_{as}/d\sigma_{bs}\sim3\times10^{-3}$. Then we find from Eq.
(\ref{Delta_nu_bc}) the corresponding $b\rightarrow c$ transition frequency shift
\begin{equation}\label{Delta_nu_bc_G}
\frac{d\Delta\nu_{bc}}{d\sigma_{bs}} =
\frac{\hbar}{ml}(a_s-a_t)\frac{G_{2s}}{K_{abs}}\sim-1.2\times10^{-7}\textrm{Hz}\cdot\textrm{cm}^2
\end{equation}
This is an order of magnitude less than the experimentally observed slope of
$\Delta\nu_{ab}(\sigma_{bs})$. Note, that the extrapolation of the 3D experimental
data~\cite{Ahokas_3D} to the 2D case within the scaling approach $n_a\rightarrow\sigma_a/l$ gives
essentially the same value as Eq.(\ref{Delta_nu_bc_G}).

It is also remarkable that the part of $\sigma_{as}$ independent of $\sigma_{bs}$ (see Eq.
(\ref{as})) contributes to the apparent wall shift $\Delta A_w(0)$ given by the extrapolation of
the experimental $\Delta\nu_{ab}(\sigma_{bs})$ to zero density. Assuming $G_{1s}\sim G_1$ this
contribution may be estimated as
\begin{equation}\label{dA_a}
\frac{\delta
A_w}{h}\simeq\left(\frac{\gamma_p}{\gamma_e}\right)\frac{2\hbar}{ml}(a_s-a_t)\frac{G_{1s}}{K_{abs}}\sim-420\,\textrm{Hz}.
\end{equation}

We should emphasize that, in contrast to the $b\rightarrow c$ transition, the $b\rightarrow a$
resonance is unaffected by the presence of residual $a$ atoms, as follows from the general
statement of Zwierlein {\it et al.}~\cite{Zwierlein}.
%{\it In fact, under the microwave-induced
%coherent rotation of (in this case) nuclear spins, the total number of nuclear triplets and
%singlets in the $a-b$ mixture does not change: the spins within the initially $a$ and $b$-state
%groups remain parallel, while the spins of atoms from different groups remain antiparallel.}

So far, we assumed the wall shift $\Delta A_w$ in Eq. (\ref{Delta_nu_fin}) to be density
independent. However, there are indications that this may not be true. In particular, Morrow {\it
et al.}~\cite{morrow} and Jochemsen {\it et al.}~\cite{jochemsen} measured the wall shift in zero
field on the surface of $^4$He and $^3$He and found $\Delta A_w$($^4$He)$=-49(2)$ kHz and $\Delta
A_w$($^3$He)$=-23(2)$ kHz. This was attributed to a wider surface profile of $^3$He as compared to
$^4$He, which results in a wider and more shallow minimum of the adsorption potential for H atoms.
Consequently, the wavefunction of adsorbed atoms spans farther from the surface, the average
distance between hydrogen and helium atoms is increased and the electric polarization of hydrogen,
which is responsible for the wall shift, is reduced.

Remarkably, the behavior of $\Delta A_w$ very closely resembles that of the binding energy $E_a$
of hydrogen to the helium surface, namely, $E_a=1.14$~\cite{morrow,OurEa} and 0.40
K~\cite{jochemsen} for $^4$He and $^3$He, respectively. Moreover, the binding energy of hydrogen
linearly decreases with an increase in the occupation of the surface state of $^3$He~\cite{OurEa}.
A respective 6\% decrease in $\Delta A_w$ was observed by Ahokas {\it et al.}~\cite{Ahokas} upon
the addition of $^3$He such that $E_a$ should decrease by $\sim10\%$. The above similarity has a
deep grounding: according to the second-order perturbation theory, a fractional change in the
unpaired electron density at the proton is
$\Delta\vert\psi_e(0)\vert^2/\vert\psi_e(0)\vert^2\simeq-2V/E_{\rm H}$, where $E_{\rm H}$ is the
average energy for the excited states of H and $V$ is the interaction energy~\cite{Adrian}, which
is seemingly the binding energy $E_a$ in our case. Thus, we may intuitively write
\begin{equation}\label{dA-dE}
\frac{\delta(\Delta A_w)}{\Delta A_w}\sim\frac{\delta E_a}{E_a}.
\end{equation}

Interaction of adsorbed H atoms with each other also changes $E_a$. In a thermal cloud of $b$
atoms ($g^{(2)}=2$),
\begin{equation}\label{dE}
\delta E_a=E_\textrm{int}=-2\sigma_b\tilde{U_t},
\end{equation}
where $\tilde{U_t}\simeq4\pi\hbar^2a_t/ml\simeq5\times10^{-15}$ K\,cm$^2$ is the effective vertex
of the triplet interaction in 2D~\cite{KSS87}. Combining Eqs. (\ref{dA-dE}) and (\ref{dE}) we
finally obtain the estimate
\begin{equation}\label{dA}
\frac{\delta(\Delta A_w)}{\Delta
A_w}\sim-\frac{2\sigma_b\tilde{U_t}}{E_a}\simeq-10^{-14}\textrm{cm}^2\times\sigma_b.
\end{equation}
For the apparent $b\rightarrow a$ frequency shift this gives
\begin{equation}\label{Delta_nu-teo}
\Delta\nu_{ab}\sim2.3\times10^{-10}\textrm{Hz}\cdot\textrm{cm}^2\times\sigma_b,
\end{equation}
about six times less than the experimental value of $C_1$.

%The estimate (\ref{dA-dE}) is also supported by our numerical evaluation a the density-dependent
%change in $A$ based on the variation of $A$ with the distance between H and He atoms~\cite{Meyer},
%and the solution of Gross-Pitaevskii equation for H atoms in the adsorption potential. The details
%of the calculation will be published elsewhere.

To our knowledge, this is the first mentioning of the density-dependent wall shift of the
hyperfine constant. This shift is associated with pair interaction. Therefore, like the contact
shift in alkalies~\cite{Harber}, it must decrease by a factor of two when the 2D hydrogen
undergoes the 2D analogue of Bose-Einstein condensation, i.e., becomes locally
coherent~\cite{Kagan2000}.

Strictly speaking, the H-H collisions may also contribute to a {\it negative} shift of $A$, by the
analogy with the hyperfine pressure shift of H atoms in a helium atmosphere~\cite{Hardy80}.
Unfortunately, the experimental data on the H-H pressure shift are unavailable. Theoretical
calculation of the effect is a formidable problem even in free space, let alone the present 2D
case (see, e.g., similar calculations for hydrogen in a helium buffer gas~\cite{Meyer}).

Combining Eqs. (\ref{Delta_nu_fin}), (\ref{Delta_nu_bc_G}) and (\ref{Delta_nu-teo}) we finally
obtain that the effect of residual $a$ atoms on $\nu_{bc}$ and the density-dependent wall shift of
$\nu_{ab}$ together amount to about 30\% of the frequency shift observed in
experiment~\cite{Ahokas}. We regard this as a qualitative agreement in view of a rather large
uncertainty in the value of the two-body relaxation rate constant $G_2$, the approximate character
of Eq. (\ref{dA-dE}) and the absence of experimental data on the effective vertex $\tilde U$. In
addition, the H-H interaction in the adsorbed phase may differ from the one characterized by the
quoted value of $\tilde U$ owing to substrate-mediated effects~\cite{Hazzard}. Obviously, the
measurement of the surface density of $a$ atoms would provide a direct check of the two
contributions to the apparent hyperfine frequency shift, as discussed in this work.

%\begin{acknowledgments}
%We are grateful to... This work was supported by...
%\end{acknowledgments}

\bibliography{Shift_rev1}

\end{document}